\documentclass[referee,epjc3]{svjour3}

\usepackage{epsfig}
\usepackage{graphicx}
\usepackage{amsfonts,amssymb}
\usepackage{color}

  \usepackage{slashed}
  \usepackage{url}
\usepackage{bm}
\usepackage{verbatim}
\usepackage{float}

\def\beq{\begin{equation}}
\def\eeq{\end{equation}}
\def\bea{\begin{eqnarray}}
\def\eea{\end{eqnarray}}
\def\ba{\begin{array}}
\def\ea{\end{array}}

\def\nn{\nonumber}

\def\Ga{\Gamma}

\def\la{\lambda}

\def\phc{\phi_c}

\def\rd{{\rm d}}

\def\rd{{\rm d}}
\def\ri{{\rm i}}

\journalname{Eur. Phys. J. C}
\begin{document}


\title{Scalar field theory in the strong self-interaction limit}


\author{Marco Frasca\thanksref{e1,addr1}}
\thankstext{e1}{e-mail: marcofrasca@mclink.it}
\institute{Via Erasmo Gattamelata, 3 \\
             00176 Roma (Italy) \label{addr1}}


\date{\today}

\maketitle

\begin{abstract}
Standard Model with a classical conformal invariance holds the promise to give a better understanding of the hierarchy problem and could pave the way for beyond the standard model physics. So, we give here a mathematical treatment of a massless quartic scalar field theory with a strong self-coupling both classically and for quantum field theory. We use a set of classical solutions recently found and show that there exists an infinite set of infrared trivial scalar theories with a mass gap. Free particles have superimposed a harmonic oscillator set of states. The classical solution is displayed through a current expansion and the next-to-leading order quantum correction is provided. Application to the Standard Model would entail the existence of higher excited states of the Higgs particle and reduced decay rates to WW and ZZ that could be already measured.
\end{abstract}

\PACS{14.80.bn, 14.80.Ec, 12.60.Fr, 11.10.Lm}



\section{Introduction}

Scalar field theory is an essential tool to master the main techniques in quantum field theory (see e.g. \cite{Peskin:1995ev,Nair:2005iw,Srednicki:2007qs}). It appeared just like a mathematical object until quite recently at LHC the Higgs particle was observed displaying all the expected properties for a scalar field interacting with other matter in the Standard Model \cite{Aad:2012tfa,Chatrchyan:2012ufa}.

Higgs field, as proposed in the sixties~\cite{Englert:1964et,Higgs:1964ia,Higgs:1964pj,Guralnik:1964eu,Higgs:1966ev,Kibble:1967sv,prl_19_1264}, is characterized by a mass term with a ``wrong'' sign and a weak quartic term providing self-interaction. The original formulation of the Standard Model postulates that conformal invariance must hold for all other matter \cite{prl_19_1264,sm_salam} that is, all particles entering into the model are massless and only breaking the symmetry SU(2)$\otimes$U(1) through the Higgs mechanism yields the mass terms. Higgs mechanism considers a potential term the same as the one in the Landau theory of phase transitions. This forces the choice of a odd mass term. The introduction of such term is the reason of the so-called ``hierarchy'' problem as the next-to-leading order correction to the mass of the Higgs field goes like the square of a cut-off running it to a Planck mass where the model is expected to fail. Just a proper fine tuning or some other mechanism yet to be discovered can explain the observed mass of this particle.

In order to evade this problem, Bardeen \cite{Bardeen:1995kv} proposed that the Standard Model should preserve conformal invariance at the classical level. This would imply that the breaking of the symmetry should be dynamical generated, possibly through radiative corrections through the  Coleman-Weinberg \cite{Coleman:1973jx} mechanism. In a recent paper \cite{Meissner:2006zh}, Nicolai and Meissner pointed out that, due to the smallness of the mass of the Higgs particle obtained using the Coleman-Weinberg mechanism, another Higgs particle must be introduced reconciling in this way Bardeen's approach with observational data. But the success of the Coleman-Weinberg mechanism, being perturbative in origin, implies that, in order to obtain the right mass, one cannot stop to the first few terms of a perturbation series. This has been recently proved by 
Chishtie, Hanif, Jia, Mann, McKeon, Sherry and Steele \cite{Chishtie:2010ni} and Steele and Wang \cite{Steele:2012av} that, extending to higher orders the computation of the effective potential, the right mass for the Higgs particle is recovered giving a boost to the idea of conformal invariance for the Standard Model. This moves the test of this idea from the existence of a further Higgs particle to the experimental determination of the self-coupling of the Higgs field. This is something to be seen at the restart of the LHC on 2015.

The aim of this paper is to show how a consistent quantum field field theory can be built assuming the self-coupling of the field large and the field itself is massless. This is obtained by using a set of exact classical solutions that were recently obtained \cite{Frasca:2009bc}. These solutions display massive nonlinear waves notwithstanding the theory is massless. An immediate consequences of this is that there exists an infinite set of quantum field theories having a trivial infrared fixed point and that have a non-null vacuum expectation value mimicking the behavior of the Higgs field as currently appears in the Standard Model. One of the immediate consequences is that higher excited states exist for the particle and that production rates for decay to WW and ZZ are different from those expected in the Standard Model, paving the way to check conformal invariance earlier from the already collected data at LHC. Similarly, we completely define the perturbative solutions in a strong self-coupled scalar theory both classically and for quantum field theory. We just note that the criticism put forward in \cite{Antipin:2013exa} is here overcome as we can have classical conformal invariance, keep it at a quantum level with dimensional regularization and maintain a physical value for the mass of the Higgs particle as we will see. This makes this scenario an even stronger competitor for a conformal extension of the Standard Model.

The paper is structured as follows. In Sec.~\ref{sec1} we introduce the classical theory and we solve it completely with a finite but not so small coupling. The Green function is also obtained that will be fundamental for the quantum analysis.In Sec.~\ref{sec:cesce}, we prove that our current expansion is really a strong coupling expansion obtaining a power series in the inverse powers of the coupling. In Sec.~\ref{sec2} we provide the current expansion and relative n-point functions that can be defined in this way for the classical solution. In Sec.~\ref{sec3} we give a quantum treatment. Firstly, we solve numerically the Dyson-Schwinger equation for the scalar field to prove consistency for our approach. Then, in the limit of strong coupling, we compute the next-to-leading order term both for the classical solution and the Green function. It is obtained an expansion in inverse powers of the coupling. In Sec.~\ref{sec4} we present the Callan-Symanzik equation and the beta function for the theory to the next-to-leading order. The field renormalization constant is computed. In Sec.~\ref{sec4a} we show how renormalization can be systematically done also for this formulation of the perturbation series in quantum field theory. In Sec.\ref{sec4b} we discuss in depth the question of how this scalar field theory breaks conformal invariance and the relevance of the existence of a zero mode. In Sec.~\ref{sec5} we comment about application of these results to the Standard Model and the production rates with respect to the Standard Model are given. Finally, in Sec.~\ref{sec6} we yield the conclusions.

\section{Classical scalar field theory}
\label{sec1}

We consider a classical scalar field $\phi$ satisfying the equation
\begin{equation}
\label{eq:eq1}
   \partial^2\phi+\lambda\phi^3=j
\end{equation}
being $\lambda>0$ the (dimensionless) strength of the self-interaction and $j$ an external source. Our aim is to get an expansion in terms of the inverse of some positive power of $\lambda$. In order to get the right perturbation series, we rescale the space-time vector as $x^\mu\rightarrow \sqrt{\lambda}x^\mu$. In the same way, we explicit the dependence on $\lambda$ of the source as $j\rightarrow\sqrt{\lambda}j$. The interesting point to note here is that this choice, that is somewhat arbitrary, fixes the expansion parameter of the perturbation series. With this choice on the current, we take
\begin{equation}
\label{eq:scalar}
  \phi(x)=\sum_{n=0}^\infty\lambda^{-\frac{n}{2}}\phi_n(x).
\end{equation}
Then, it is not difficult to see that the following set of equations holds
\begin{eqnarray}
   \partial^2\phi_0+\phi_0^3&=& 0 \nonumber \\
   \partial^2\phi_1+3\phi_0^2\phi_1&=&j \nonumber \\
   \partial^2\phi_2+3\phi_0^2\phi_2&=& -3\phi_0\phi_1^2 \nonumber \\
   \partial^2\phi_3+3\phi_0^2\phi_3&=&-6\phi_0\phi_1\phi_2-\phi_1^3 \nonumber \\
   \partial^2\phi_4+3\phi_0^2\phi_4&=&-3\phi_0\phi_2^2-3\phi_1^2\phi_2-6\phi_0\phi_1\phi_3 \nonumber \\
   &\vdots.&
\end{eqnarray}
From this set of equations, we recognize that we have essentially a couple of equations to solve before to solve completely the theory in the limit we are interested in. We have to find the exact solution to the following system of equations:
\begin{eqnarray}
   \partial^2\varphi(x)+\varphi^3(x)&=&0 \nonumber \\
   \partial^2G(x,x')+3\varphi^2(x)G(x,x')&=&\delta^4(x-x').
\end{eqnarray}
$G(x,x')$ is a fundamental solution. This approach is quite general provided we are able to get $G(x,x')$. Indeed, a set of exact solutions exist for these two equations. The solution for the first one is \cite{Frasca:2009bc}
\begin{equation}
\label{eq:exact}
   \varphi(x)=\mu2^\frac{1}{4}{\rm sn}(k\cdot x+\theta,-1)
\end{equation}
provided that
\begin{equation}
    k^2=\frac{1}{\sqrt{2}}\mu^2
\end{equation}
being $\mu$ and $\theta$ two integration constants and sn a Jacobi elliptic function. This represents a kind of massive solution even if we started from a massless field theory when $k$ is interpreted like the momenta of the wave. This class of solutions has the property, similarly to the case of the plane waves of the free theory, to have finite energy density \cite{MO131417}. This reduces the second equation to
\begin{equation}
\label{eq:green}
   \partial^2G(x,x')+3\mu^22^\frac{1}{2}{\rm sn}^2(k\cdot x+\theta,-1)G(x,x')=\delta^4(x-x').
\end{equation}
We immediately notice that the Green function is not translation invariant. This equation is linear and we use a gradient expansion to solve it. We must remember that we are working with distributions and their derivatives. In the following sections we will show that: {\em Green function is indeed translational invariant} and that the {\em strong coupling expansion is equivalent to an expansion in the powers of current}.

\subsection{Green function}
\label{subsec:green}

Let us rewrite eq.(\ref{eq:green}) as
\begin{equation}
   \partial_t^2G(x,x')+3\mu^22^\frac{1}{2}{\rm sn}^2(k\cdot x+\theta,-1)G(x,x')=\delta^4(x)+\epsilon\Delta_2G(x,x')
\end{equation}
where we have introduced an arbitrary order parameter $\epsilon$ that we set to 1 at the end of computation. So, taking $G(x,x')=\sum_{n=0}^\infty\epsilon^nG_n(x,x')$, we get the set of equations
\begin{eqnarray}
   \partial_t^2G_0(x,x')+3\mu^22^\frac{1}{2}{\rm sn}^2(k\cdot x+\theta,-1)G_0(x,x')&=&\delta^4(x-x') \nonumber \\
   \partial_t^2G_1(x,x')+3\mu^22^\frac{1}{2}{\rm sn}^2(k\cdot x+\theta,-1)G_1(x,x')&=&\Delta_2G_0(x,x') \nonumber \\
   \partial_t^2G_2(x,x')+3\mu^22^\frac{1}{2}{\rm sn}^2(k\cdot x+\theta,-1)G_2(x,x')&=&\Delta_2G_1(x,x') \nonumber \\
   &\vdots.&
\end{eqnarray}
By noting that $G_0(x,x')=\delta^3(x-x')\bar G(t,t')$, the leading order reduces to solve the equation, that corresponds to the original equation but in the rest frame,
\begin{equation}
\label{eq:exGF}
   \partial_t^2\bar G(t,t')+3\mu^22^\frac{1}{2}{\rm sn}^2(k_0t+\theta,-1)\bar G(t,t')=\delta(t-t')
\end{equation}
being $k_0=\mu/2^\frac{1}{4}$. We can yield an exact solution in the form
\begin{equation}
\label{eq:AppA}
   G_0(t,0,{\bm x},{\bm x}') = -\delta^3(x-x')\frac{1}{\mu 2^\frac{3}{4}}H(t){\rm dn}\left(\frac{\mu}{2^\frac{1}{4}}t+\theta,-1\right){\rm cn}\left(\frac{\mu}{2^\frac{1}{4}}t+\theta,-1\right)
\end{equation}
being $H(t)$ the Heaviside step function and provided we fix the phases to $\theta=(4m+1)K(-1)$ with $m=0,1,2,\ldots$. $K(\alpha)=\int_0^\frac{\pi}{2}dy/\sqrt{1-\alpha\sin^2y}$ is the complete elliptic integral of the first kind. But an exact solution can be also provided for $G(t,t')$ \cite{klei} as we will show in \ref{sec:AppA}. In this way we have identified an infinite  set of solutions to the classical scalar field theory and for these solutions the corresponding quantum theory is trivial \cite{Frasca:2009bc} at the leading order.  We are able to solve exactly eq.~(\ref{eq:green}). Indeed, we have the formal series
\begin{equation}
\label{eq:series}
    G(t,0,{\bm x}-{\bm x}')=G_0(t,0,{\bm x}-{\bm x}')+\int dt'G_0(t,t',{\bm x}-{\bm x}')\Delta_2\delta^3(x)
		+\int dt'dt''G_0(t,t',{\bm x}-{\bm x}')G_0(t',t'',{\bm x}-{\bm x}')\Delta_2\delta^3(x)\Delta_2\delta^3(x)+\ldots
\end{equation}
that can be easily resummed using a Fourier transform and with the fact that $\int dt'G_0(t,t',{\bm x}-{\bm x}')=G_0(t,0,{\bm x}-{\bm x}')$. Firstly, we note that $({\rm sn}(x,-1))'={\rm cn}(x,-1){\rm dn}(x,-1)$ and so
\begin{equation}
    G_0(t,0,{\bm x}-{\bm x}')=\delta^3(x-x')\bar G(t,0)=
		-\delta^3(x-x')\frac{1}{\mu 2^\frac{3}{4}}H(t)\left.\frac{d}{du}{\rm sn}(u,-1)
		\right|_{u=\frac{\mu}{2^\frac{1}{4}}t+\theta}.
\end{equation}
But one has 
\begin{equation}
    {\rm sn}(u,-1)=\frac{2\pi}{K(-1)}\sum_{n=0}^\infty(-1)^n\frac{e^{-\left(n+\frac{1}{2}\right)\pi}}{1+e^{-(2n+1)\pi}}
    \sin\left((2n+1)\frac{\pi}{2K(-1)}u\right)
\end{equation}
and this gives
\begin{eqnarray}
   G_0(t,0,{\bm x}-{\bm x}')&=&-\delta^3(x-x')\frac{1}{\mu 2^\frac{3}{4}}H(t)\frac{\pi^2}{K^2(-1)}\sum_{n=0}^\infty(-1)^n(2n+1)
   \frac{e^{-\left(n+\frac{1}{2}\right)\pi}}{1+e^{-(2n+1)\pi}}\times \nonumber \\
   &&\cos\left((2n+1)\frac{\pi}{2K(-1)}\frac{\mu}{2^\frac{1}{4}}t+(2n+1)(4m+1)\frac{\pi}{2}\right).
\end{eqnarray}
In the following we choose the simplest realization for $m=0$ and so,
\begin{equation}
   G_0(t,0,{\bm x}-{\bm x}')=-\delta^3(x-x')\frac{1}{\mu 2^\frac{3}{4}}H(t)\frac{\pi^2}{K^2(-1)}\sum_{n=0}^\infty(2n+1)
   \frac{e^{-\left(n+\frac{1}{2}\right)\pi}}{1+e^{-(2n+1)\pi}}
   \sin\left((2n+1)\frac{\pi}{2K(-1)}\frac{\mu}{2^\frac{1}{4}}t\right).
\end{equation}
We note that our solution must be invariant under time reversal $t\rightarrow -t$ as also the time reversed solution must be kept into account. This means that our solution is
\begin{equation}
   G_0(t,0,{\bm x}-{\bm x}')=\delta^3(x-x')[\bar G(t,0)+\bar G(-t,0)].
\end{equation}
This will provide us the Fourier transformed result
\begin{equation}
   G_0(p_0,0)=\sum_{n=0}^\infty\frac{B_n}{p_0^2-m_n^2+i\epsilon}
\end{equation}
where we put
\begin{equation}
    B_n=(2n+1)^2\frac{\pi^3}{4K^3(-1)}\frac{e^{-(n+\frac{1}{2})\pi}}{1+e^{-(2n+1)\pi}}.
\end{equation}
and $m_n=(2n+1)\frac{\pi}{2K(-1)}\left(\frac{1}{2}\right)^\frac{1}{4}\mu$. Turning to our solution series eq.~(\ref{eq:series}) we recognize that higher order terms are just the geometric series that adds a ${\bm p}^2$ to the denominator granting for Lorentz invariance. So, the final result is
\begin{equation}
\label{eq:prop}
   G_0(p)=\sum_{n=0}^\infty\frac{B_n}{p^2-m_n^2+i\epsilon}.
\end{equation}
In this way we were able to recover the translation invariance of the theory we started from. It is interesting to note that $\sum_nB_n=1$ and the theory recovers the free limit for $\lambda=0$ and so $m_n=0$.

It is essential to notice that the full propagator we have got, eq.~(\ref{eq:prop}), {\sl is translationally invariant} and this is proved {\sl a posteriori}. This can also be seen by demanding Lorentz invariance to the solution of the theory. In Sec.~\ref{sec3.1} we will show numerically that eq.~(\ref{eq:prop}) solves the Dyson-Schwinger equations for the scalar field.

\subsection{Strong coupling solution}

We are now in a position to provide a strong coupling solution for eq.~(\ref{eq:scalar}) using the set of perturbation equations just obtained. We will get (omitting the homogeneous solutions as usual in this case)
\begin{eqnarray}
\label{eq:sols}
   \phi_1(x)&=&\int d^4x_1 G(x,x_1)j(x_1) \nonumber \\
   \phi_2(x)&=&-3\int d^4x_1G(x,x_1)\phi_0(x_1)\left[\int d^4x_2G(x_1,x_2)j(x_2)\right]^2 \nonumber \\
   \phi_3(x)&=&18\int d^4x_1G(x,x_1)\phi_0(x_1)\int d^4x_2G(x_1,x_2)j(x_2)\int d^4x_3G(x_1,x_3)\phi_0(x_3)\times \nonumber \\
   &&\left[\int d^4x_4G(x_3,x_4)j(x_4)\right]^2-\int d^4x_1G(x,x_1)\left[\int d^4x_2G(x_1,x_2)j(x_2)\right]^3 \nonumber \\
   \phi_4(x)&=&-27\int d^4x_1G(x,x_1)\phi_0(x_1)
   \left\{\int d^4x_2G(x_1,x_2)\phi_0(x_2)\left[\int d^4x_3G(x_2,x_3)j(x_3)\right]^2\right\}^2+ \nonumber \\
   &&9\int d^4x_1G(x,x_1)\left[\int d^4x_2 G(x_1,x_2)j(x_2)\right]^2
   \int d^4x_3G(x_1,x_3)\phi_0(x_3)\times \nonumber \\
   &&\left[\int d^4x_4G(x_3,x_4)j(x_4)\right]^2- \nonumber \\
   &&108\int d^4x_1G(x,x_1)\phi_0(x_1)\int d^4x_2 G(x_1,x_2)j(x_2)
   \int d^4x_3G(x_1,x_3)\phi_0(x_3)\times \nonumber \\
   &&\int d^4x_4G(x_3,x_4)j(x_4)\int d^4x_5G(x_4,x_5)\phi_0(x_5)\times \nonumber \\
   &&\left[\int d^4x_6G(x_5,x_6)j(x_6)\right]^2+\nonumber \\
   &&6\int d^4x_1G(x,x_1)\phi_0(x_1)\int d^4x_2 G(x_1,x_2)j(x_2)\int d^4x_3G(x_1,x_3)\times \nonumber \\
   &&\left[\int d^4x_4G(x_3,x_4)j(x_4)\right]^3 \nonumber \\
   &\vdots.&
\end{eqnarray}
We recognize here a series expansion into power of currents that is, being $\phi=\phi[j]$, we have
\begin{eqnarray}
\label{eq:currents}
   \phi[j]&=&\phi[0]+\int d^4x_1\left.\frac{\delta\phi}{\delta j(x_1)}\right|_{j=0}j(x_1)+ \nonumber \\
   &&\frac{1}{2!}\int d^4x_1d^4x_2\left.\frac{\delta^2\phi}{\delta j(x_1)\delta j(x_2)}\right|_{j=0}j(x_1)j(x_2)+ \nonumber \\
   &&\frac{1}{3!}\int d^4x_1d^4x_2d^4x_3\left.\frac{\delta^3\phi}{\delta j(x_1)\delta j(x_2)\delta j(x_3)}\right|_{j=0}j(x_1)j(x_2)j(x_3)+ \nonumber \\
   &&\frac{1}{4!}\int d^4x_1d^4x_2d^4x_3d^4x_4
   \left.\frac{\delta^4\phi}{\delta j(x_1)\delta j(x_2)\delta j(x_3)\delta j(x_4)}\right|_{j=0}j(x_1)j(x_2)j(x_3)j(x_4)+ \nonumber \\
   &&\ldots.
\end{eqnarray}
So, this completes our proof that our strong coupling expansion is equivalent to a current expansion for the solution of eq.~(\ref{eq:eq1}).

\section{Current expansion is a strong coupling expansion}
\label{sec:cesce}

In this section we will show that, despite the rescaling of space-time variables and current, we have got indeed a strong coupling expansion. We assume whatever value of the coupling $\lambda$, avoid any rescaling and just write eq.~(\ref{eq:eq1}) as
\begin{equation}
   \partial^2\phi+\lambda\phi^3=\epsilon j
\end{equation}
being $\epsilon$ an arbitrary parameter we use in our perturbation series just as a bookkeeper and we will set it at 1 at the end of computation. The exact solution is now
\begin{equation}
\label{eq:exact2}
   \varphi(x)=\mu\left(\frac{2}{\lambda}\right)^\frac{1}{4}{\rm sn}(k\cdot x+\theta,-1)
\end{equation}
holding the condition
\begin{equation}
    k^2=\sqrt{\frac{\lambda}{2}}\mu^2.
\end{equation}
Now, introducing the series $\phi=\sum_{n=0}^\infty\epsilon^n\phi_n$ we recover the set of equations
\begin{eqnarray}
   \partial^2\phi_0+\lambda\phi_0^3&=& 0 \nonumber \\
   \partial^2\phi_1+3\lambda\phi_0^2\phi_1&=&j \nonumber \\
   \partial^2\phi_2+3\lambda\phi_0^2\phi_2&=& -3\lambda\phi_0\phi_1^2 \nonumber \\
   \partial^2\phi_3+3\lambda\phi_0^2\phi_3&=&-6\lambda\phi_0\phi_1\phi_2-\lambda\phi_1^3 \nonumber \\
   \partial^2\phi_4+3\lambda\phi_0^2\phi_4&=&-3\lambda\phi_0\phi_2^2-3\lambda\phi_1^2\phi_2-6\lambda\phi_0\phi_1\phi_3 \nonumber \\
   &\vdots.&
\end{eqnarray}
that is identical to the set obtained with the rescaling in $\lambda$ that we must keep here to prove our assertion. So, the solution of this set of equation is rather straightforward and yields
\begin{eqnarray}
\label{eq:sols2}
   \phi_1(x)&=&\int d^4x_1 G(x,x_1)j(x_1) \nonumber \\
   \phi_2(x)&=&-3\lambda\int d^4x_1G(x,x_1)\phi_0(x_1)\phi_1^2(x_1)\nonumber \\
   \phi_3(x)&=&-6\lambda\int d^4x_1G(x,x_1)\phi_0(x_1)\phi_1(x_1)\phi_2(x_1)-3\lambda\int d^4x_1G(x,x_1)\phi_1^3(x_1) \nonumber \\
   &\vdots&
\end{eqnarray}
that we stop at the third order being useless to go to higher orders for our aims. I order to give a proof of our assertion, that this is indeed a strong coupling expansion, we consider the low-energy limit of the propagator of the theory (\ref{eq:prop}). This reduces to a contact term due to the presence of the mass gap, that is,
\begin{equation}
   G(x,x')=-\delta^4(x-x')\sum_{n=0}^\infty\frac{B_n}{m_n^2}=-\frac{1}{\mu^2\sqrt{\lambda}}c_0\delta^4(x-x')
\end{equation}
being $c_0$ a numerical constant, as the series converges, that is not important now. We just notice the dependence on the inverse of the square root of $\lambda$. So, one has
\begin{eqnarray}
   \phi_1(x)&=&-\frac{c_0}{\mu^2\sqrt{\lambda}}j(x) \nonumber \\
   \phi_2(x)&=&\frac{3c_0^3}{\mu^4\lambda^\frac{3}{4}}\tilde\phi_0(x)j^2(x)\nonumber \\
   \phi_3(x)&=&-\frac{c_0^4}{\mu^6\lambda}\left[18c_0\tilde\phi_0^2(x)+3\right]j^3(x) \nonumber \\
   &\vdots.&
\end{eqnarray}
We see that our current expansion goes like $\sum_{n=1}^\infty\lambda^{-(n+1)/4}a_n(x)j^n(x)$ and this completes our proof: {\em A current expansion is a strong coupling expansion}. The choice of the energy range cannot overturn the asymptotic nature of this series and so, our conclusion is a general one.

\section{Classical n-point functions and higher order corrections}
\label{sec2}

Given the current expansion in eq.~(\ref{eq:currents}), we can identify a set of n-point functions for the classical field theory. This can be easily achieved by comparing the series in eq.~(\ref{eq:currents}) with the results obtained into eq.~(\ref{eq:sols}). As usual we take $\phi=\phi[j]$ and we get by computing the functional derivatives
\begin{eqnarray}
   G_2(x,x_1)&=&\left.\frac{\delta\phi}{\delta j(x_1)}\right|_{j=0}=G_0(x,x_1) \nonumber \\
   G_3(x,x_1,x_2)&=&\frac{1}{2!}\left.\frac{\delta^2\phi}{\delta j(x_1)\delta j(x_2)}\right|_{j=0}=
   -3\int d^4x_3G_0(x,x_3)\phi_0(x_3)G_0(x_3,x_1)G_0(x_3,x_2) \nonumber \\
   G_4(x,x_1,x_2,x_3)&=&
   \frac{1}{3!}\left.\frac{\delta^3\phi}{\delta j(x_1)\delta j(x_2)\delta j(x_3)}\right|_{j=0}= \nonumber \\
   &&18\int d^4x_4d^4x_5G_0(x,x_4)\phi_0(x_4)G_0(x_4,x_1)G_0(x_4,x_5)\times \nonumber \\
   &&\phi_0(x_5)G_0(x_5,x_2)G_0(x_5,x_3)- \nonumber \\
   &&\int d^4x_4G_0(x,x_4)G_0(x_4,x_1)G_0(x_4,x_2)G_0(x_4,x_3) \nonumber \\
   G_5(x,x_1,x_2,x_3,x_4)&=&
   \frac{1}{4!}\left.\frac{\delta^4\phi}{\delta j(x_1)\delta j(x_2)\delta j(x_3)\delta j(x_4)}\right|_{j=0}= \nonumber \\
   &&-27\int d^4x_5G_0(x,x_5)\phi_0(x_5)\int d^4x_6G_0(x_5,x_6)\phi_0(x_6)\times \nonumber \\
   &&\int d^4x_7G_0(x_1,x_7)\phi_0(x_7)G_0(x_6,x_1)G_0(x_6,x_2)G_0(x_7,x_3)G_0(x_7,x_4)+ \nonumber \\
   &&9\int d^4x_5G_0(x,x_5)G_0(x_5,x_1)G_0(x_5,x_2)\times \nonumber \\
   &&\int d^4x_6G_0(x_5,x_6)\phi_0(x_6)G_0(x_6,x_3)G_0(x_6,x_4)- \nonumber \\
   &&108\int d^4x_5G_0(x,x_5)\phi_0(x_5)G_0(x_5,x_1)\int d^4x_6G_0(x_5,x_6)\phi_0(x_6)\times \nonumber \\
   &&G_0(x_6,x_2)\int d^4x_5G_0(x_2,x_5)\phi_0(x_5)G_0(x_5,x_3)G_0(x_5,x_4)+ \nonumber \\
   &&6\int d^4x_5G_0(x,x_5)\phi_0(x_5)G_0(x_5,x_1)\int d^4x_6G_0(x_5,x_6)\times \nonumber \\
   &&G_0(x_6,x_2)G_0(x_6,x_3)G_0(x_6,x_4) \nonumber \\
   &\vdots&.
\end{eqnarray}
In this way we have given explicitly the n-point functions till $n=5$. This shows that eq.~(\ref{eq:currents}) agrees perfectly well with eq.~(\ref{eq:sols}) as expected but we have given it using the Green function $G_0(x,x')$ just computed. These integrals could need regularization even if we are working in the classical case.

\section{Quantum corrections}
\label{sec3}

\subsection{Numerical Dyson-Schwinger equations and Green function}
\label{sec3.1}

Numerical solution of Dyson-Schwinger equations is a standard tool to check for an analytical solution for n-point functions to be correct. This entails some kind of truncation to the infinite tower of such equations and the evaluation of angular integration. In this section we will follow the approach pointed out in \cite{Aguilar:2004sw} for Yang-Mills equations. Dyson-Schwinger equations for the scalar field are well-known \cite{Swanson:2010pw}. We just write them here for our case with a given classical solution as we have discussed above. One has
\begin{eqnarray}
\Delta^{-1}(p^2) &=& G_0^{-1}(p^2) +  3i\lambda \int \frac{d^dq}{(2\pi)^d} \Delta(q^2) +  \nonumber\\
&& i\lambda\int \frac{d^d\ell_1}{(2\pi)^d}\, \frac{d^d\ell_2}{(2\pi)^d} \, \frac{d^d \ell_3}{(2\pi)^d} \Delta(\ell_1^2) \Delta(\ell_2^2) \Delta(\ell_3^2) \, \Gamma(\ell_1,\ell_2,\ell_3,-p)\times \nonumber \\
&& (2\pi)^d \delta(\ell_1+\ell_2+\ell_3-p).
\label{eq:DS}
\end{eqnarray}
To treat numerically this equation we need to move to the Euclidean space, fix the vertex to $\Gamma = -i6\lambda$ (rainbow-ladder approximation) and evaluate the angular part of the integrals. In the Euclidean space and for the given vertex this equation becomes
\begin{eqnarray}
\Delta^{-1}(p^2) &=& -G_0^{-1}(p^2) +  3\lambda \int \frac{d^dq}{(2\pi)^d} \Delta(q^2) +  \nonumber\\
&& -6\lambda^2\int \frac{d^d\ell_1}{(2\pi)^d}\, \frac{d^d\ell_2}{(2\pi)^d} \Delta(\ell_1^2) \Delta(\ell_2^2) \Delta((p-\ell_1-\ell_2)^2).
\label{eq:DSe}
\end{eqnarray}
For the evaluation of the angular part of the last integral we work in the same way as in \cite{Aguilar:2004sw}. This is a further approximation we introduced: For an integration variable $q$ and the external momenta $p$, when $q^2 < p^2$ we take $\Delta((q-p)^2)\rightarrow\Delta(p^2)$ and when $q^2 > p^2$ we can set $\Delta((q-p)^2)\rightarrow \Delta(q^2)$. We are able to prove that our propagator (\ref{eq:prop}) numerically solves the Dyson-Schwinger equation when a constant $\mu$ for the mass spectrum exists such that the error between the numerical solution and the exact equation (\ref{eq:prop}) can be reduced at less than 1\%.  The solution is obtained by iteration starting with a test function chosen to be a Yukawa free propagator. The solution is seen to converge toward eq.~(\ref{eq:prop}) after very few iterations.

\begin{figure}[H]
\begin{center}
\includegraphics[angle=0, width=\textwidth]{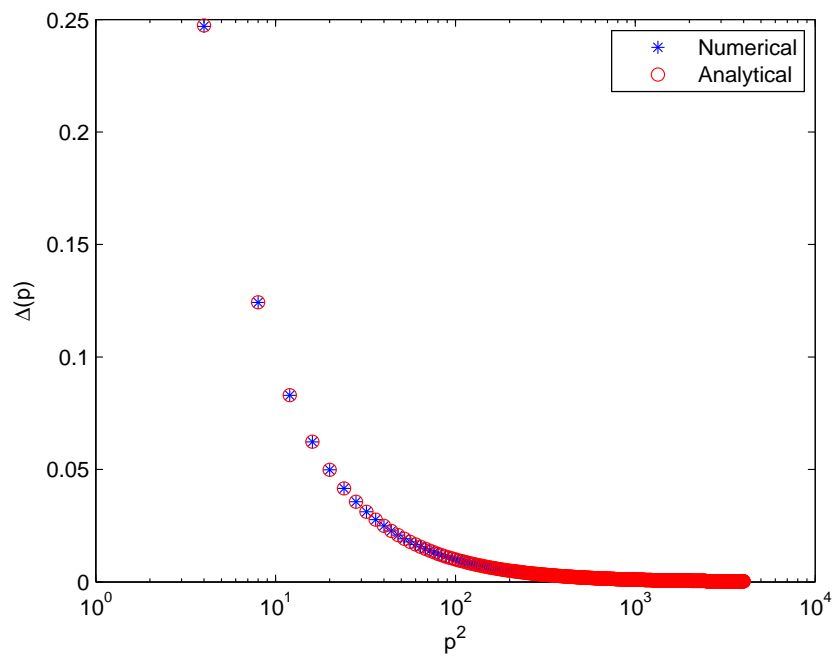}
\caption{\label{fig:fig0} Numerical solution to the truncated Dyson-Schwinger equation compared to the exact solution (\ref{eq:prop}) at very small $\lambda$ when momenta overcome the mass term.The agreement is excellent as it should.}
\end{center}
\end{figure}

To be sure of the consistency of our approximations, we checked that in the limit of a very small $\lambda$ the expected free solution $1/p^2$ is recovered. This must be so also for eq.~(\ref{eq:prop}) as, when mass terms are negligible with respect to momenta then $\sum_nB_n=1$ and the massless free solution is the one expected. This is shown in Fig.~\ref{fig:fig0}.

\begin{figure}[H]
\begin{center}
\includegraphics[angle=0, width=\textwidth]{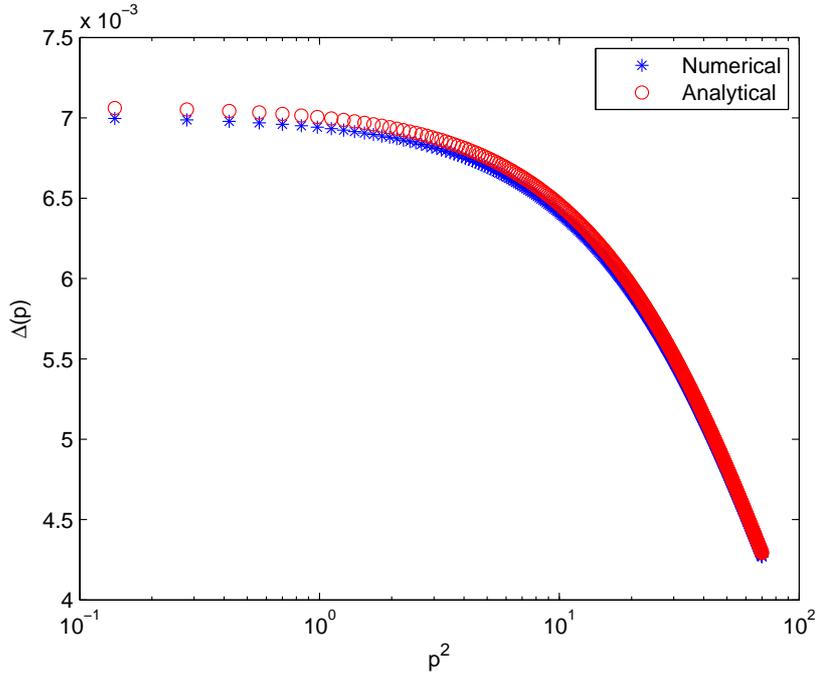}
\caption{\label{fig:fig1} Numerical solution to the truncated Dyson-Schwinger equation compared to the exact solution (\ref{eq:prop}) in dimensionless units. $\lambda=1$, $\mu=10$ (dimensionless units). We have stopped the analysis with an error lesser than 1\%.}
\end{center}
\end{figure}

The case at increasing $\lambda$ is given in Fig.~\ref{fig:fig1}. We obtained this by varying properly $\mu$ in eq.~(\ref{eq:prop}) until the error becomes very small.
This proves that such a constant $\mu$ exists and that eq.~(\ref{eq:prop}) gives a correct representation in the quantum theory.

\subsection{Next-to-leading order correction}

Now, we do quantum field theory in the same limit of $\lambda\rightarrow\infty$. Let us consider the generating functional
\begin{equation}
   Z[j]=\int[d\phi]\exp\left[i\int d^4x\left(\frac{1}{2}(\partial\phi)^2-\frac{\lambda}{4}\phi^4+j\phi\right)\right].
\end{equation}
As already done in the classical case, we rescale $x\rightarrow\sqrt{\lambda}x$ and $j\rightarrow \sqrt{\lambda}j$. So, we can rewrite
\begin{equation}
   Z[j]=\int[d\phi]\exp\left[\frac{i}{\lambda}\int d^4x
   \left(\frac{1}{2}(\partial\phi)^2-\frac{1}{4}\phi^4+\frac{1}{\sqrt{\lambda}}j\phi\right)\right].
\end{equation}
Then, we use the analytic solution (\ref{eq:exact}) by taking the exact identity $\phi=\phi_0+\frac{1}{\sqrt{\lambda}}\delta\phi$ amounting to a simple shift for the integration variable. This gives
\begin{equation}
   Z[j]={\cal N}e^{\frac{i}{\lambda\sqrt{\lambda}}\int d^4xj\phi_0}\int[d\delta\phi]e^{\frac{i}{\lambda^2}\int d^4x
   \left[\frac{1}{2}(\partial\delta\phi)^2-\frac{3}{2}\phi_0^2\delta\phi^2+j\delta\phi\right]}
   e^{-\frac{i}{\lambda^2}\int d^4x\left(\frac{1}{\sqrt{\lambda}}\phi_0\delta\phi^3+\frac{1}{4\lambda}\delta\phi^4\right)}
\end{equation}
where use has been made of the equation of motion $\partial^2\phi_0+\phi_0^3=0$. We are in a position to do perturbation theory using the Green function given in eq.~(\ref{eq:prop}). It is interesting to note that this theory has a non-null value on the vacuum. This can be easily seen from the first exponential factor and noting also that $\phi_0(0)=\mu(2/\lambda)^\frac{1}{4}\ne 0$ where we have reinserted the coupling constant $\lambda$. What we have classically are nonlinear oscillations around this constant value and this explains why the excitations of the theory are massive notwithstanding we started from a massless theory. Then, it is easy to write down this generating functional for perturbation theory \cite{Nair:2005iw}
\begin{equation}
  Z[j]={\cal N}e^{i\sqrt{\lambda}\int d^4xj\phi_0}
  e^{-i\int d^4x\left(-\frac{1}{\sqrt{\lambda}}\phi_0(x)\frac{\delta^3}{i\delta j(x)^3}+\frac{1}{4\lambda}\frac{\delta^4}{\delta j(x)^4}\right)}
  e^{\frac{i}{2}\int d^4xd^4yj(x)G_0(x-y)j(y)}
\end{equation}
where we have undone the space-time scaling at this stage. This completes our formulation of a quantum scalar field theory with a strong self-interaction and we are able to do perturbation theory in the inverse of the coupling. We note an odd contribution for quantum corrections to the classical solution and an even one with a well-known form but with a quite different propagator. This propagator is meaningful in the infrared having a finite limit for $p\rightarrow 0$ going like $1/\sqrt{\lambda}$. Now, we rewrite the above functional in a more manageable form \cite{Nair:2005iw}
\begin{equation}
   Z[j]={\cal N}e^{i\sqrt{\lambda}\int d^4xj\phi_0}e^{\frac{i}{2}\int d^4xd^4yj(x)G_0(x-y)j(y)}
   \left[e^{\int d^4xd^4yj(x)G_0(x-y)\frac{\delta}{\delta(\delta\phi(y))}}{\cal F}[\phi]\right]_{\delta\phi=0}
\end{equation}
being
\begin{equation}
   {\cal F}[\phi]=\exp\left[-\frac{i}{2}\int d^4xd^4y\frac{\delta}{\delta(\delta\phi(x))}G_0(x-y)\frac{\delta}{\delta(\delta\phi(y))}\right]
   \exp\left[-i\int d^4x\left(\frac{1}{\sqrt{\lambda}}\phi_0\delta\phi^3+\frac{1}{4\lambda}\delta\phi^4\right)\right]
\end{equation}

Now, we go on by computing the next to leading order correction. One has,
\begin{eqnarray}
   Z[j] &=& {\cal N}e^{i\sqrt{\lambda}\int d^4xj\phi_0}e^{\frac{i}{2}\int d^4xd^4yj(x)G_0(x-y)j(y)}\times \nonumber \\
   &&\left(1-\frac{3}{2\sqrt{\lambda}}G_0(0)\int d^4xd^4yj(x)G_0(x-y)\phi_0(y)+\right. \nonumber \\
   &&\left.\frac{3i}{2\lambda}G_0(0)\int d^4xd^4yd^4zj(x)G_0(x-y)G_0(z-y)j(z)+\ldots\right).
\end{eqnarray}
where use has been made of the equation $G_0(x-z)=\int d^4yG_0(x-y)G_0(y-z)$. We can complete this computation by evaluating $G_0(0)$. In order to perform this evaluation, we just note that the theory has a natural energy scale, $\mu$, to be used as a cut-off. So, we want to compute
\begin{equation}
    G_0(0)=\sum_{n=0}^\infty B_n\int \frac{d^4p}{(2\pi)^4}\frac{1}{p^2-m_n^2+i\epsilon}.
\end{equation}
This integral, if we imply the limit $\lambda\rightarrow\infty$ at the end of computation and use the physical cut-off arising from the classical solutions
(we will discuss this choice in the next section),
can be evaluated exactly to give
\begin{equation}
    \left.G_0(0)\right|_{\Lambda^2=\mu^2}=\frac{\mu^2}{16\pi^2}-\frac{1}{16\pi^2}\sum_{n=0}^\infty B_nm_n^2\ln\left(\frac{m_n^2}{m_n^2-\mu^2}\right)
\end{equation}
where use has been made of the equation $\sum_{n=0}^\infty B_n=1$. Now, $\mu$ is finite, being a physical constant in the infrared limit, and so the formal limit $\lambda\rightarrow\infty$ produces $0$ for the sum in the second term of rhs. This result comes out to be the same as seen in weak perturbation theory due to the structure of the propagator in the infrared that is a sum of Yukawa propagators that have identical structure to the ultraviolet case. For both limits the theory is trivial.

Now, we can evaluate the next-to-leading order correction to the classical solution from quantum field theory. We get
\begin{equation}
   \frac{1}{i\sqrt{\lambda}}\left.\frac{\delta Z[j]}{\delta j(x)}\right|_{j=0}=
   \phi_0(x)-\frac{3\mu^2}{32\pi^2\lambda}\int d^4y[-iG_0(x-y)]\phi_0(y)+\ldots.
\end{equation}
Similarly, the two-point function just gives
\begin{equation}
     \frac{1}{i^2\lambda}\left.\frac{\delta^2 Z[j]}{\delta j(x)\delta j(y)}\right|_{j=0}=
     -iG_0(x-y)+\frac{3i\mu^2}{16\pi^2\lambda^2}\int d^4zG_0(x-z)G_0(y-z)+\ldots.
\end{equation}
that can be Fourier transformed into
\begin{equation}
\label{eq:delta}
     i\Delta(p^2) = G_0(p^2)-\frac{3\mu^2}{16\pi^2\lambda^2}[G_0(p^2)]^2+\ldots.
\end{equation}
In both equations we undid the current normalization through $\sqrt{\lambda}$. We just note that higher order corrections to the propagator can also depend on $\phi_0$. Here we have got the term that renormalize the masses $m_n$.

\subsection{Duality principle}

There is an interesting relation between this formalism in the strong coupling limit and the standard weak coupling expansion yielding the quadratic correction to the mass of the Higgs field. This is the duality principle in perturbation theory firstly formulated in \cite{Frasca:1998ch}. This applies both to the classical and quantum theory. This principle states that to have a strong or weak coupling expansion depends just on the choice of the perturbation. So, let us consider our case
\begin{equation}
   \partial^2\phi+\lambda\phi^3=j.
\end{equation}
One can choose $\lambda\phi^3$ as a perturbation and will get a weak coupling expansion $\phi=\sum_n\lambda^n\tilde\phi_n$. Choosing the current $j$ as a perturbation instead will yield a strong coupling expansion $\phi=\sum_n\lambda^{-n}\hat\phi_n$ (but see Sec.~\ref{sec:cesce}) that we properly identified as a current expansion. In order to understand better how this can come about, we divide the equation by $\lambda$ and obtain
\begin{equation}
   \lambda^{-1}\partial^2\phi+\phi^3=\lambda^{-1}j.
\end{equation}
A meaning can only be attached to it if we now assume $\lambda\rightarrow\infty$ and recognize that the perturbation is the current $j$ instead. In this way this becomes a boundary layer problem in perturbation theory as we have a small parameter multiplying the derivative term in the differential equation. So, interchanging the perturbation term into the equation gives perturbation series with an expansion parameter one the inverse of the other. This is the essence of the duality principle in perturbation theory.

This is a general property of differential equations that we applied to the case of the scalar field theory and can be extended to quantum field theory in the way we displayed in this paper. Using this principle it is possible to study a theory in almost all the range of variability of its parameters.

\section{Callan-Symanzik equation}
\label{sec4}

The leading order propagator represents the one of a free theory, according to K\"allen-Lehman representation. In the infrared limit, the free particles of the theory have a superimposed harmonic oscillator spectrum. Being a free theory in the infrared limit, one should expect also that the running coupling goes to zero in this limit. This is exactly what we see using a Callan-Symanzik equation, that is
\begin{equation}
    \mu\frac{\partial G_0(p^2)}{\partial\mu}-\beta(\lambda)\frac{\partial G_0(p^2)}{\partial\lambda}+(2-2\gamma(\lambda)) G_0(p^2)=0
\end{equation}
provided that
\begin{equation}
    \beta(\lambda)=4\lambda\qquad\gamma(\lambda)=1.
\end{equation}
This beta function was already obtained by others \cite{Suslov:2010rk}.
This result immediately implies 
\begin{equation}
   \lambda_r(p)=\lambda\frac{p^4}{\Lambda^4}
\end{equation}
being $\Lambda$ a proper momenta cut-off that we will discuss in the next section. We see that, while the bare coupling can be large, the theory reaches a trivial infrared fixed point lowering momenta. On the other side, for enough large momenta, we get an increasing coupling but the theory has also a trivial ultraviolet infrared fixed point and so, there must be a maximum for the running coupling at increasing momenta. One can fix $\Lambda$ in this way.

Now, one can compute the next-to-leading order quantum corrections to the classical results. To show this we use a standard approach (see e.g. \cite{Peskin:1995ev}). From eq.~(\ref{eq:delta}) we approximate
\begin{equation}
    i\Delta(p^2)\approx\frac{G_0(p^2)}{1+\frac{3\mu^2}{16\pi^2\lambda^2}G_0(p^2)}
\end{equation}
and so
\begin{equation}
    \Delta(p^2)\approx -i\sum_{n=0}^\infty B_n\frac{1}{p^2-m_n^2-\delta m_n^2(p^2)}
\end{equation}
where we have put
\begin{equation}
    \delta m_n^2(p^2) = -\frac{3\mu^2}{16\pi^2\lambda^2}(p^2-m_n^2)G_0(p^2)
\end{equation}
and so
\begin{equation}
    \delta m_n^2(0) = -\frac{3\mu^2}{16\pi^2\lambda^2}m_n^2G_0(0)=c_0\frac{3m_n^2}{16\pi^2\lambda^\frac{5}{2}}
\end{equation}
being $c_0=0.7071067811\ldots$. This means that we have
\begin{equation}
     M^2_n(\lambda) = m^2_n\left(1+c_0\frac{3}{16\pi^2\lambda^\frac{5}{2}}+\ldots\right)
\end{equation}
that, remembering that $m^2_n\propto\sqrt{\lambda}$, can be seen as a renormalization of the coupling giving
\begin{equation}
     \lambda_R^\frac{1}{2}=\lambda^\frac{1}{2}+c_0\frac{3}{16\pi^2\lambda^2}+\ldots.
\end{equation}
This identifies also the renormalization constant for the field being defined through $M^2_n=Zm_n^2$ and so
\begin{equation}
    Z = 1+c_0\frac{3}{16\pi^2\lambda^\frac{5}{2}}+\ldots.
\end{equation}

\section{Systematic of renormalization}
\label{sec4a}

In the preceding section we have shown how to perform computations in quantum field theory when the scalar field theory is strongly coupled. In order to give soundness to such a computation we need to show how the theory can be managed in this limit in a systematic way. We do this by reducing our generating functional to that of a standard renormalizable theory. So, we undo all our scaling and rewrite the generating functional of the theory as
\begin{equation}
\label{eq:zsc}
  Z[j]={\cal N}e^{i\int d^4xj\phi_0}
  e^{-i\int d^4x\left(-\frac{1}{\sqrt{\lambda}}\phi_0(x)\frac{\delta^3}{i\delta j(x)^3}+\frac{1}{4\lambda}\frac{\delta^4}{\delta j(x)^4}\right)}
  e^{\frac{i}{2}\int d^4xd^4yj(x)G_0(x-y)j(y)}
\end{equation}
where now is $\phi_0(x)=\mu(2/\lambda)^\frac{1}{4}{\rm sn}(p\cdot x+\theta,-1)$ and $m_n=(2n+1)\frac{\pi}{2K(-1)}\left(\frac{\lambda}{2}\right)^\frac{1}{4}\mu$. For the following, it is important to notice that ${\rm sn}(u,-1)=\sum_{n=0}^\infty(-1)^n(\pi^2/2K^2(-1))e^{-(n+1/2)\pi}/(1+e^{-(2n+1)\pi})\sin((2n+1)\pi u/2K(-1))$ and so, it is just a sum of exponentials that contribute to the conservation of momenta in a vertex. Now, we show that this theory is renormalizable exactly in the same way as it is done in the weak perturbation case. So, let us consider the scalar theory
\begin{equation}
     {\cal L}=\frac{1}{2}(\partial\phi)^2-g_1\phi^3-\frac{g_2}{4}\phi^4.
\end{equation}
This theory is known to be renormalizable in $d=4$ having $g_1$ dimension 1 and being $g_2$ dimensionless (e.g. see \cite{Peskin:1995ev}) and all the cut-off dependencies can be reabsorbed into $g_1$, $g_2$ and the field. The cubic term is super-renormalizable in $d=4$ and it is not relevant in the ultraviolet. Writing down the generating functional, this takes the form
\begin{equation}
\label{eq:sc34}
  Z'[j]={\cal N}
  e^{-i\int d^4x\left(g_1\frac{\delta^3}{i\delta j(x)^3}+\frac{g_2}{4}\frac{\delta^4}{\delta j(x)^4}\right)}
  e^{\frac{i}{2}\int d^4xd^4yj(x)\hat G_0(x-y)j(y)}
\end{equation}
being $\hat G_0(x-y)$ a solution to the equation $\partial^2\hat G_0(x-y)=\delta^4(x-y)$. Feynman diagrams can be immediately written down from the generating functional (\ref{eq:sc34}). We see that this functional can be exactly mapped onto the one of the strongly coupled expansion (\ref{eq:zsc}) provided we properly account for the field $\phi_0(x)$ but this just contributes adding momentum to the vertex. Besides, the propagator of the strongly coupled theory is, in agreement with K\"allen-Lehman representation, that of a free theory (a sum of free massive propagators) as the theory has a trivial infrared fixed point.

So, we can conclude that the renormalization program applies as well to the strongly coupled perturbation theory with all the cut-off dependencies eventually reabsorbed into the coupling $\lambda$ and the field. This conclusion is important as this means that we are able to get a renormalized perturbation theory both for the coupling that goes formally to zero and to infinity. It is also interesting to note that the classical solution gets quantum corrections as it should.

The choice of the cut-off is based on the conclusions we draw from the renormalization group analysis we performed in the preceding section. This theory has a non-perturbative beta function providing a running coupling going to zero at both the extrema of the momentum range. This means that there must exist a maximum at some value of the momentum range and the existence of this upper bound for the coupling permits us to fix a proper momentum scale for the theory. In this way, we chose the same cut-off emerging from classical solutions down to all the range, being this arbitrary.


\section{Breaking of conformal symmetry}
\label{sec4b}

We are in a position to understand the kind of breaking of symmetry that occurs with these classical solutions. In this section we will follow the approach presented by Nicolai and Meissner in \cite{Meissner:2007xv} that set the framework to understand how conformal invariance is preserved granting a solution to the hierarchy problem. We will show that our approach is perfectly consistent with that presented by those authors. As shown in \cite{Meissner:2007xv}, the behavior of a theory under conformal symmetry can be understood when use is made of a energy-momentum tensor that respects such a symmetry. This was yielded in \cite{Callan:1970ze} (see also a more recent discussion in \cite{Forger:2003ut}) and for our case takes the form
\begin{equation}
    T^{\mu\nu}=\partial^\mu\phi\partial^\nu\phi-\frac{1}{2}\eta^{\mu\nu}\left(\partial^\kappa\phi\partial_\kappa\phi-\frac{\lambda}{2}\phi^4\right)
		+\frac{1}{6}\left(\eta^{\mu\nu}\partial^2-\partial^\mu\partial^\nu\right)\phi^2.
\end{equation}
The conditions for preserving conformal invariance is given by $\eta_{\mu\nu}T^{\mu\nu}=T^\mu_{\ \mu}=0$, i.e. the trace of the tensor is zero. We expect that this condition is violated by renormalization in quantum field theory. We have already shown (preceding section) that renormalization in our case is performed exactly in the same way is done for a weak coupling case. As proved in \cite{Meissner:2007xv}, one loop correction is local and goes like $\phi^4$ eventually correcting the anomalous Ward identity that is expected to be $\beta(\lambda)O_4$ with $O_4$ a fourth order operator. We are able to prove that the same argument applies to our case as well.

For the classical solution given in (\ref{eq:exact}), we note that the traceless condition holds on shell \cite{Callan:1970ze,Forger:2003ut} and, being this solution an exact one, this condition holds straightforwardly. From this we can conclude that the symmetry is spontaneously broken due to the massive excitation we get. In a moment we will show that a mode zero exists in this case that works as a Goldstone mode (this was already pointed out in \cite{Frasca:2009bc}). Firstly, we would like to point out that, for the quantum corrections, the same argument given in \cite{Meissner:2007xv} applies also to our case. For our aims, we write here the result in \cite{Meissner:2007xv} for the effective potential in the Coleman-Weinberg approach
\bea
\Ga^{(1)}[\phc] &=& \frac{3\la}2 \int d^4x
    \, D(x,x;\phc) \,\phc^2(x) \; +
    \label{gaform}\\
    && \!\!\!\!\!\!\!\!\!\!\!\!\!\!\!\!\!\!\!\!\!
   + \; \frac{9\la^2}4\int d^4x
   \int d^4y \, D(x,y;\phc) \, \phc^2(y)\,
        D(y,x;\phc) \, \phc^2(x) \; + \dots  \nn
\eea
provided that
\beq
\big(\partial^2 + 3\la \phc^2(x)\big) D(x,y;\phc)
= \delta^{(4)}(x,y).
\label{propag}
\eeq
But we know how to solve this equation and so, the exact form of the propagator for the classical solution (\ref{eq:exact}) making this identical to the case of a constant field discussed in \cite{Meissner:2007xv} when the expansion is taken around $\phi_c(x)=\phi_0(x)$. In order to complete the calculation, we have to take into account that we will have (written for the Euclidean case as customary)
\beq
D(x,y;\phc) \big|_{\phc(x)=\phi_0(x)} =
\int\frac{\rd^4 p}{(2\pi)^4}e^{-\ri p(x-y)}\sum_{n=0}^\infty\frac{B_n}{p^2+ m_n^2}
\eeq
with $B_n$ and $m_n$ given in Sec.~\ref{subsec:green}. So, dimensional regularization can be applied as well and the structure of the integrals is identical being that of sums of free massive propagators. From this result we can conclude that the problem of the quadratic divergence is similar to that discussed in the weak coupling limit.

Finally, let us discuss the question of the Goldstone mode arising from the spontaneous breaking of the conformal symmetry. We have already discussed this question in \cite{Frasca:2009bc}. We give this here for reader's convenience. Our key operator is $M=\partial^2+3\lambda\phi_0^2(x)$. We have shown in sec.\ref{sec3} that this operator quantifies the fluctuations with respect to the vacuum solution. A zero mode for this operator can give rise to a spontaneous breaking of symmetry. This is seen both in quantum field theory and statistical mechanics as well \cite{Bender:1992yd,Kim:2003zya,Daviaud:2013phf}. For our aims it is enough to show that the kernel of the $M$ operator is not trivial. Indeed, the eigenvalue problem takes the form
\begin{equation}
   [\partial^2+3\lambda\phi_0^2(x)]\chi_n(x)=\lambda_n\chi_n(x).
\end{equation}
It is really easy to see that the case with $\lambda_n=0$ admits a non-trivial solution \cite{Frasca:2009bc,Frasca:2013gba} and so, a zero mode exists. This implies that $(\det M)^{-1}$ is infinite.


\section{Higgs model}
\label{sec5}

This analysis of a scalar field theory appears well suited to application to the Standard Model in the conformal limit. Indeed, it appears not distinguishable from a Higgs field but decay rates are modified. This can be immediately realized if we look at the propagator of the theory that can be easily interpreted through K\"allen-Lehman representation as the sum of an infinite number of states each one having mass $m_n$ and a probability of production $B_n^2$. This factor, being lesser than 1, can depress decay rates of processes like $H\rightarrow WW,\ ZZ$ \cite{Frasca:2013ada} that are currently observed at LHC. We just point out that the number of events obtained so far by ATLAS and CMS is too small yet to rule out this model. We give here a table of these probabilities for each excited state to give a correct view of what one should expect \cite{Frasca:2013ada}.
\begin{table}[!ht]
\begin{center}
\begin{tabular}{|c|c|c|c|} \hline\hline
n & $B_n^2$        & \%    & \% to SM \\ \hline
0 & 0.6854746582   &  -    &   31     \\ \hline 
1 & 0.2780967321   & 59    &   72     \\ \hline
2 & 0.0333850484   & 95    &   97     \\ \hline
3 & 0.0028276899   & 99.6  &   99.7   \\ \hline
4 & 0.0002019967   & 99.97 &   99.98  \\ \hline
\end{tabular}
\caption{\label{tab:fn} Weights and percentage reductions of the decay rates of Higgs excited states. Percentages are respect to the ground state in the second column and respect to the Standard Model (SM) in the third one.}
\end{center}
\end{table}
It is also important to note that higher massive states, if ever exist, are increasingly difficult to observe due to the even more depressed production rates with respect to the ground state, that should be the currently observed Higgs particle at LHC, as can be evinced from Tab.~\ref{tab:fn}.

It is interesting to digress on the ground state, that is the currently seen Higgs particle, that has a decay factor of $\mu\approx 0.68$. This means that, with respect to the expectations of the Standard Model, the production rates for the decays $H\rightarrow ZZ$ and $H\rightarrow WW$ are reduced. Indeed, CMS data reach almost exactly this value for the WW decay but errors are sizable yet. Rates are $\mu=0.93\pm 0.27$ for ZZ and $\mu=0.72\pm 0.19$ for WW. These figures are in really close agreement with those we computed in Tab.\ref{tab:fn} but, as said, error bars are too much large to make a claim. For ATLAS, results are better aligned with the Standard Model but a similar argument on the size of errors holds and in order to get a definite answer we will have to wait the restart of the LHC.

\section{Conclusions}
\label{sec6}

We have shown how a massless scalar field theory with a quartic self-interaction can be properly managed in the strong coupling limit. The theory yields massive excitations notwithstanding no mass term is present. This would permit to build up a fully conformal Standard Model, at a classical level, and agrees with recent results using Coleman-Weinberg mechanism \cite{Steele:2012av}. We would expect that, if one is able to resum all the radiative corrections, in the end, our result should be recovered. This appears quite difficult, at the present, but our approach is already amenable to experimental tests. However, a recent computation of higher order corrections to Coleman-Weinberg mechanism points toward a peculiar structure of singularity of the complete effective potential that could be a precursor to further excited states \cite{Hanif:2013npa} in agreement with our approach.

In the end, even if this kind of mechanism should not be observed, it is nevertheless interesting the fact that a perturbation theory for a strongly coupled scalar field can be developed much in the same way this happens for weak perturbations. 

\section*{Acknowledgments}
I would like to thank the Editors Gino Isidori and Keith Olive for their help in the review process. The referee has been instrumental in the improvement of the paper.

\appendix
\section{Exact Green function}
\label{sec:AppA}

Our aim is to give the exact solution for the Green function in eq.~(\ref{eq:exGF}). The equation to solve is (dot is the derivative with respect to time)
\begin{equation}
\label{eq:AexGF}
   \ddot {\bar G}(t,t')+3\kappa^2{\rm sn}^2(\kappa t/\sqrt{2}+\theta,-1){\bar G}(t,t')=\delta(t-t')
\end{equation}
with $\kappa^2=\mu^2\sqrt{2\lambda}$. We have inserted back all the constants for the sake of clearness. To solve this equation we use the technique described in \cite{klei}. This can be applied by noticing that we know two independent solutions of the equation
\begin{equation}
\label{eq:AexGF2}
   \partial_t^2\Delta(t,t')+3\kappa^2{\rm sn}^2(\kappa t/\sqrt{2}+\theta,-1)\Delta(t,t')=0.
\end{equation}
These are
\begin{equation}
   y_1(t)={\rm cn}(\kappa t/\sqrt{2}+\theta,-1){\rm dn}(\kappa t/\sqrt{2}+\theta,-1)
\end{equation}
and the other can be obtained writing it as
\begin{equation}
   y_2(t)=y_1(t)\cdot w(t)
\end{equation}
with
\begin{equation}
   {\rm cn}(\kappa t/\sqrt{2}+\theta,-1){\rm dn}(\kappa t/\sqrt{2}+\theta,-1)\ddot w
	-2\sqrt{2}\kappa{\rm sn}^3(\kappa t/\sqrt{2}+\theta,-1)\dot w=0.
\end{equation}
One has
\begin{equation}
    y_2(t)=\frac{\sqrt{2}}{4}\kappa t{\rm cn}(\kappa t/\sqrt{2}+\theta,-1){\rm dn}(\kappa t/\sqrt{2}+\theta,-1)
		+\frac{1}{4}{\rm sn}(\kappa t/\sqrt{2}+\theta,-1).
\end{equation}
We can write \cite{klei}
\begin{equation}
    \Delta(t,t')=\alpha(t')y_1(t)+\beta(t')y_1(t)w(t)
\end{equation}
and we have to require
\begin{equation}
    \Delta(t',t')=0\qquad \left.\dot\Delta(t,t')\right|_{t=t'}=-1
\end{equation}
yielding a set of equations to get $\alpha(t')$ and $\beta(t')$. The final result is ($\theta=0$)
\begin{eqnarray}
    G(t,t')&=&\frac{1}{2\kappa}H(t-t')
		\left(\kappa(t-t')
		{\rm cn}(\kappa t/\sqrt{2},-1)
		{\rm dn}(\kappa t/\sqrt{2},-1)
		{\rm cn}(\kappa t'/\sqrt{2},-1)
		{\rm dn}(\kappa t'/\sqrt{2},-1)\right. \nonumber \\
		&&+\sqrt{2}
		{\rm sn}(\kappa t/\sqrt{2},-1)
		{\rm cn}(\kappa t'/\sqrt{2},-1)
		{\rm dn}(\kappa t'/\sqrt{2},-1) \nonumber \\
		&&\left.-\sqrt{2}
		{\rm sn}(\kappa t'/\sqrt{2},-1)
		{\rm cn}(\kappa t/\sqrt{2},-1)
		{\rm dn}(\kappa t/\sqrt{2},-1)
		\right).
\end{eqnarray}
We recover the Green function yielded in the article, eq.~(\ref{eq:AppA}), choosing $t'$ so that ${\rm cn}(\kappa t'/\sqrt{2},-1)=0$. The Jacobi function cn is a periodic function and so there are infinite values $t'=\bar t$ that satisfies this equation. One has
\begin{equation}
    \hat G(t)=-\frac{1}{\sqrt{2}\kappa}H(t-\bar t)
		{\rm cn}(\kappa t/\sqrt{2},-1)
		{\rm dn}(\kappa t/\sqrt{2},-1),
\end{equation}
and for $t'=0$ it is
\begin{eqnarray}
    \hat G(t,0)&=&\frac{1}{2\kappa}H(t)
		(\kappa t
		{\rm cn}(\kappa t/\sqrt{2},-1)
		{\rm dn}(\kappa t/\sqrt{2},-1)
		+\sqrt{2}
		{\rm sn}(\kappa t/\sqrt{2},-1)).
\end{eqnarray}
A similar formula was obtained in Ref.~\cite{Frasca:2005sx}. We note that both $y_1(t)$ and $y_2(t)$ can have infinite values for which are zero. This means that this theory has zero modes \cite{Frasca:2013gba} and to the Green functions an arbitrary solution of eq.~(\ref{eq:AexGF2}) can always be added.



\end{document}